\documentclass[aps,showpacs,prb,twocolumn]{revtex4}

\usepackage{graphicx}
\usepackage{dcolumn}
\usepackage{bm}

\begin{document}


\title{Effects of pressure on the ferromagnetic state of the CDW compound SmNiC$_2$}

\author{B. Woo, S. Seo, E. Park, J. H. Kim, D. Jang, T. Park}
\affiliation{Department of Physics, Sungkyunkwan University, Suwon 440-746, Korea}
\author{H. Lee, F. Ronning, J. D. Thompson}
\affiliation{Los Alamos National Laboratory, Los Alamos, New Mexico 87545, USA}
\author{V. A. Sidorov}
\affiliation{Los Alamos National Laboratory, Los Alamos, New Mexico 87545, USA\\
Institute for High Pressure Physics of Russian Academy of Sciences, RU-142190 Troitsk, Moscow, Russia\\
Moscow Institute of Physics and Technology, RU-141700 Dolgoprudny, Moscow Region, Russia}
\author{Y. S. Kwon}
\affiliation{Department of Emerging Materials Science, Daegu Gyeongbuk Institute of Science and Technology (DGIST), Daegu 711-873, Korea}


\begin{abstract}
We report the pressure response of charge-density-wave (CDW) and ferromagnetic (FM) phases of the rare-earth intermetallic SmNiC$_2$ up to 5.5~GPa. The CDW transition temperature ($T_{CDW}$), which is reflected as a sharp inflection in the electrical resistivity, is almost independent of pressure up to 2.18~GPa but is strongly enhanced at higher pressures, increasing from 155.7~K at 2.2~GPa to 279.3~K at 5.5~GPa. Commensurate with the sharp increase in $T_{CDW}$, the first-order FM phase transition, which decreases with applied pressure, bifurcates into the upper ($T_{M1}$) and lower ($T_c$) phase transitions and the lower transition changes its nature to second order above 2.18~GPa. Enhancement both in the residual resistivity and the Fermi-liquid $T^2$ coefficient $A$ near 3.8~GPa suggests abundant magnetic quantum fluctuations that arise from the possible presence of a FM quantum critical point.
\end{abstract}
\pacs{71.27.+a, 68.35.Rh, 71.45.Lr, 75.50.Cc}
\maketitle

Low dimensional metallic systems have attracted much interest because of their propensity towards an ordered phase. Density waves are prominent examples in quasi-one-dimensional compounds, where a large anisotropy in the Fermi surface leads to a structural instability accompanied by a periodic lattice distortion\cite{gruner94}. Sensitivity to the Fermi surface topology makes it relatively easy to tune the ordered phases via such external parameters as chemical doping, pressure, and magnetic fields. For the transition-metal dichalcogenide TiSe$_2$, a CDW transition temperature is suppressed with increasing Cu intercalation and is intercepted by a dome of superconductivity centered around a projected critical concentration where the extrapolated $T_{CDW}$ becomes zero.\cite{morosan06} External pressure acts similarly to suppress the CDW phase of TiSe$_2$, inducing superconductivity in the vicinity of a projected CDW critical point.\cite{budko07} These results both by Cu intercalation and external pressure suggest that correlated electrons spontaneously adjust to a new emergent phase in the vicinity of a quantum critical point.

Rare-earth intermetallic compounds $Re$NiC$_2$ ($Re=$La, Ce, Nd, Sm, Gd, Tb, Er) show various ground states of CDWs and magnetism.\cite{onodera98, murase04} Among the intermetallics, SmNiC$_2$ is unique in that it becomes ferromagnetic, while other members are prone to an antiferromagnetic instability. X-ray scattering studies of SmNiC$_2$ reveal satellite peaks corresponding to an incommensurate wave vector (0.5,~0.52,~0) below 148~K, signaling formation of a charge-density wave.\cite{shimomura09} The abrupt disappearance of the satellite peak at the ferromagnetic transition temperature ($T_c=17.4$~K) indicates a destruction of the CDW phase and a strong correlation between the FM and CDW phases. First-principles electronic structure calculations find that Fermi-surface nesting is important for the CDW state and weaker nesting in the ferromagnetic phase leads to the destruction of the CDW below $T_c$~(ref.~7). Kim et al. recently estimated that hydrostatic pressure will enhance the CDW because the lattice constant of the Ni chain along the a-axis decreases faster than other axes, thus enhancing the Fermi surface nesting quality.\cite{kim13} Here, we report the electrical resistivity of SmNiC$_2$ under pressure up to 5.5~GPa. The CDW transition at 151.7~K (=$T_{CDW}$) is almost independent of pressure up to 2.18~GPa but linearly increases thereafter to 279.4~K at 5.5~GPa. Commensurate with the change in the CDW phase, an additional CDW phase observed at 1.47~GPa and 75~K (=$T_{CDW2}$) initially increases with pressure, reaches a peak at 2.18~GPa and is suppressed with further increasing pressure. The first-order ferromagnetic transition at 17.4~K that completely replaces the CDW state decreases with pressure and changes its nature to second order above 2.18~GPa, the critical pressure where $T_{CDW}$ sharply increases while the pressure-induced $T_{CDW2}$ starts to decrease. Even though enhancement both in the residual resistivity and the Fermi-liquid $T^2$ coefficient $A$ near 3.8~GPa suggests abundant magnetic quantum fluctuations, appearance of new magnetic phases at 3.8 and 5.4~GPa hides the possible presence of a FM quantum critical point.

SmNiC$_2$ polycrystals were synthesized by arc melting.\cite{kim12} The constituent elements of Sm, Ni, and C were prepared at a 1.1:1:2 molecular weight ratio because Sm has higher vapor pressure. Polycrystals synthesized in a tetra arc furnace were annealed at 900 $^{\circ}$C for 10~days. X-ray powder diffraction showed that they form in a single phase with the CeNiC$_2$-type orthorhombic crystalline structure and with lattice constants $a=3.7073 \AA$, $b=4.5294 \AA$, and $c=6.0998 \AA$. Pressure measurements to 5.5~GPa were performed by using a toroid-type anvil cell with an alumina-epoxy gasket and a glycerol-water mixture as a pressure medium inside the gasket. The superconducting transition temperature of lead was used to determine the pressure in the cell.\cite{eiling81, sidorov05} The electrical resistivity of SmNiC$_2$ was measured by a conventional four-probe technique via an LR700 Resistance Bridge from 300~K to 1.2~K in a $^4$He cryostat.
\begin{figure}[tbp]
\includegraphics[width=7.5cm]{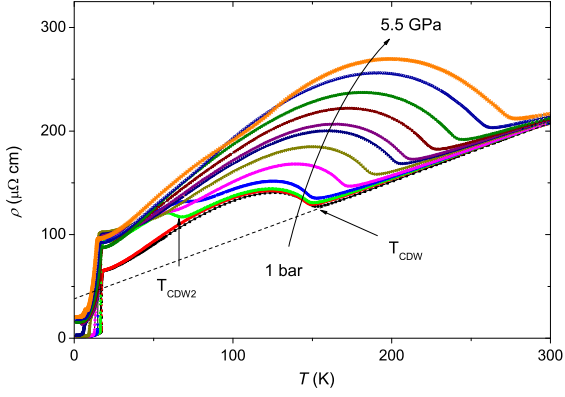}
\caption{(Color online) Temperature dependence of the electrical resistivity of SmNiC$_2$ for pressures of 1~bar, 0.87, 1.47, 2.18, 2.72, 3.22, 3.70, 3.86, 4.28, 4.7, 5.14, and 5.5~GPa from the tail to the head of the arrow. $T_{CDW}$ denotes a CDW (charge-density-wave) phase transition temperature, while $T_{CDW2}$ is a pressure-induced CDW transition temperature at a lower temperature, which appears at pressures above 1.47~GPa (green line). The dashed line is a guide to eyes that shows a linear-in-\textit{T} dependence of the high-temperature resistivity at ambient pressure.} 
\label{fig1}
\end{figure}

Figure~1 shows the electrical resistivity $\rho$ of SmNiC$_2$ as a function of temperature under pressure. At ambient pressure, $\rho$ decreases linearly with decreasing temperature and shows an inflection due to a gap opening below the CDW transition temperature $T_{CDW}$ (=151.7~K). With further decreasing temperature, $\rho$ decreases by an order of magnitude due to electrical conduction in ungapped portions of the Fermi surface and the destruction of the CDW gap at the ferromagnetic transition temperature $T_c$ (=17.4~K). The low-temperature resistivity follows a $T^2$ Landau-Fermi liquid behavior with $\rho_0=1.798~\mu\Omega\cdot$cm, where the large residual resistivity ratio (RRR=115) indicates high quality of the specimen. The Sommerfeld coefficient $\gamma$ estimated from the coefficient $A$ (=$6.08\times10^{-4}~\mu\Omega\cdot$cm$\cdot$K$^{-2}$) and the Kadowaki-Woods ratio\cite{kadowaki86} ($R_{KW}=A/\gamma^2=1.0\times10^-5~\mu\Omega\cdot$mol$^2\cdot$cm$\cdot$K$^{-2}\cdot$mJ$^{-2}$) is 7.80~mJ$\cdot$mol$^{-1}\cdot$K$^{-2}$, which is similar to $\gamma$ (=8~mJ$\cdot$mol$^{-1}\cdot$K$^{-2}$) obtained from specific heat measurements.\cite{kim12}

\begin{figure}[tbp]
\includegraphics[width=7.5cm]{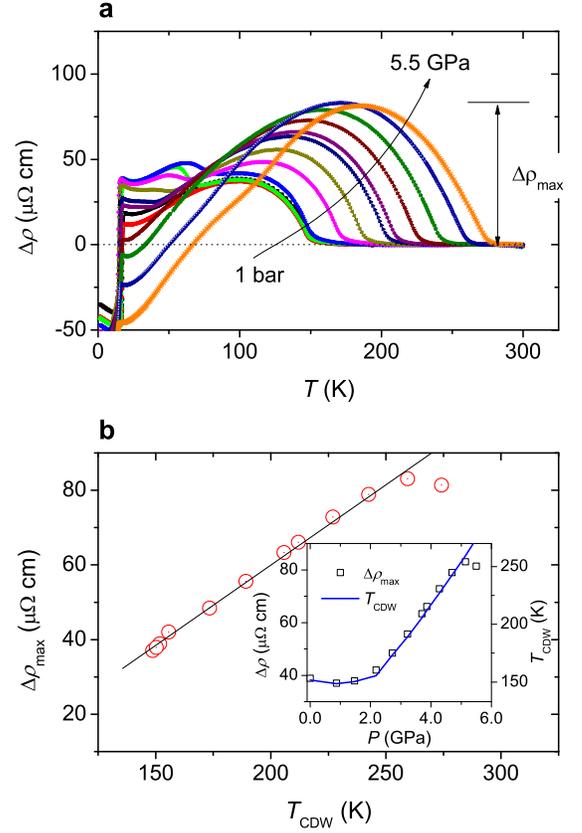}
\caption{(Color online) (a) Electrical resistivity difference of SmNiC$_2$ as a function of temperature for 1~bar, 0.87, 1.47, 2.18, 2.72, 3.22, 3.70, 3.86, 4.28, 4.7, 5.14, and 5.5~GPa from the tail to the head of the arrow, where the resistivity difference $\Delta \rho$ is obtained by subtracting the high-$T$ linear behavior: $\Delta \rho = \rho - (a+bT)$. (b) Maximum of the resistivity difference ($\Delta \rho_{max}$) plotted as a function of the CDW transition temperature $T_{CDW}$. The solid line is a guide to eyes and reflects a linear proportionality between the two parameters. Inset: Pressure dependence of $\Delta \rho_{max}$ (open squares) and $T_{CDW}$ (solid line) is plotted on the left and right ordinates, respectively.}
\end{figure}

Figure~2a depicts the resistivity difference under pressure, $\Delta \rho = \rho - (a+bT)$, where a linear background contribution observed at higher temperatures (see dashed line in Fig.~1 at ambient pressure) is subtracted from $\rho$ of SmNiC$_2$. At 0.87~GPa, as shown in Fig.~2b inset, $T_{CDW}$ decreases to 149.6~K at a depression rate of 2.4~K/GPa and the maximal value in the resistivity difference ($\Delta \rho_{max}$) also decreases. With further increasing pressure, however, $T_{CDW}$ reaches a minimum near 1.47~GPa and increases, showing a linear-in-$P$ dependence at higher pressures with a slope of 15~K/GPa. Figure~2b shows that $\Delta \rho _{max}$ linearly depends on $T_{CDW}$, indicating that the CDW gap opening is responsible for the charge carrier depletion and, thus, the increase in $\Delta \rho_{max}$. We note that there is a deviation from linearity at the highest pressure 5.5~GPa, which may be intrinsic and merits additional study at pressures higher than 5.5~GPa.

\begin{figure}[tbp]
\includegraphics[width=8cm]{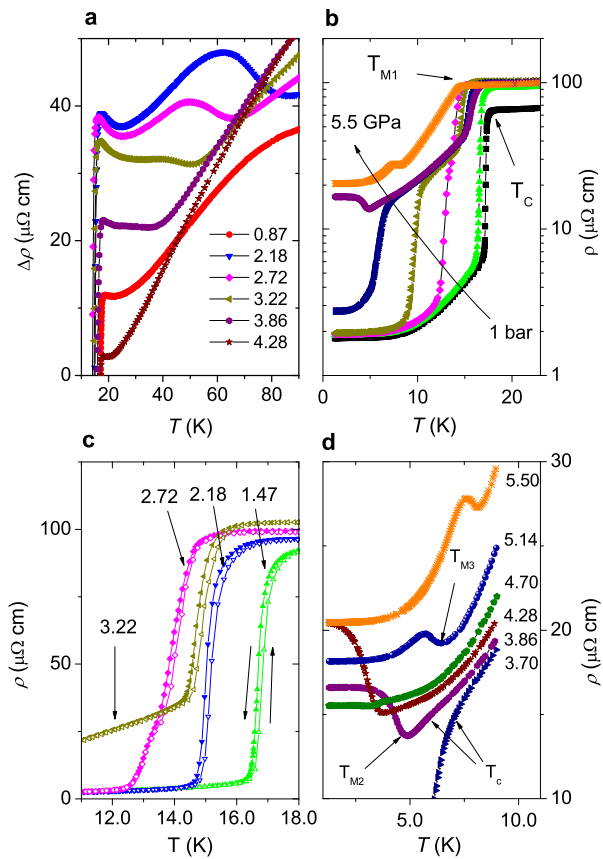}
\caption{(Color online) (a) Electrical resistivity difference of SmNiC$_2$ over a limited temperature range to show the pressure-induced CDW phase transition from 0.87 to 4.28~GPa. (b) Electrical resistivity of SmNiC$_2$ plotted on a semi-logarithmic scale near the ferromagnetic transition temperature $T_c$ for 1~bar, 1.47, 2.72, 3.22, 3.70, 3.86, and 5.5~GPa from the tail to the head of the arrow. $T_{M1}$ is assigned to a pressure-induced phase transition for $P >2.18$~GPa. (c) Thermal hysteresis of $\rho$ selectively shown for 1.47, 2.18, 2.72, and 3.22~GPa. (d) Low-temperature electrical resistivity of SmNiC$_2$ as a function of temperature plotted to demonstrate contrasting behavior across 3.8~GPa. $T_{M2}$ and $T_{M3}$ describe two emergent phase transitions for $P>3.8$~GPa. Applied pressure for each data set is indicated in units of GPa.}
\end{figure}

Thermal diffuse scattering observed by x-ray scattering measurements at ambient pressure indicates a critical phonon softening at two characteristic wavevectors of $q_1=(0.5, 0.52, 0)$ and $q_2=(0.5, 0.5, 0.5)$~(ref.~6). Only $q_1$ evolves into a CDW phase, while $q_2$ remains diffusive. A change in the Fermi surface can be incurred by applied pressure because of anisotropic elastic moduli. A slight suppression of $T_{CDW}$ at low pressures manifests a weakening of the Fermi surface nesting along $q_1$, which opens the possibility for a new competing phase. Indeed, an additional inflection in the resistivity occurs at 68.5~K and 1.47~GPa (see Fig.~2a). Considering that there already exists lattice softening along the $q_2$ wavevector, the new feature may correspond to a CDW gap opening along that direction. As shown in Fig.~3a, the CDW2 transition temperature $T_{CDW2}$ increases with pressure, reaches a maximum near 2.18~GPa, and is completely suppressed above 3.86~GPa. Here the transition temperature $T_{CDW2}$ was determined from the point of inflection. The optimal pressure for CDW2 ($\approx 2.18$~GPa) coincides with the critical pressure where the original $T_{CDW}$ starts to increase sharply, underlying that the Fermi surface topology is important to the multiple CDW phases of SmNiC$_2$. 

Figure~3b shows the resistivity of SmNiC$_2$ under pressure. The first-order ferromagnetic transition temperature $T_c$, below which Sm moments with 0.32~$\mu_B$ are aligned parallel to the a-axis,\cite{onodera98} is gradually suppressed with pressure. At pressures higher than 2.18~GPa, a plateau appears in the FM transition region, indicating that the FM ground state is accessed through an intermediate phase. The initial drop in $\rho$ is assigned as $T_{M1}$, while the second drop at a lower temperature as $T_c$ because the resistivity value approaches that of ambient pressure. These results suggest that the FM transition temperature decreases continuously, but the intermediate phase observed at a higher transition temperature $T_{M1}$ shows a dome shape with maximal value near 4.28~GPa (see Fig.~4a). Similar to Er$_5$Ir$_4$Si$_{10}$, where a CDW and antiferromagnetism coexists,\cite{galli02} it is likely that the $T_{M1}$ and CDW phases of SmNiC$_2$ coexist for $T_c \leq T \leq T_{M1}$, while the CDW disappears below $T_c$. Commensurate with the appearance of the intermediate $M1$ phase, as shown in Fig.~3c, thermal hysteresis in $\rho$ for the FM transition (or the lower phase transition) is not evident within the limit of the resistivity measurements, indicating a second order or a weakly first order nature for pressures above 2.18~GPa.

\begin{figure}[tbp]
\includegraphics[width=7cm]{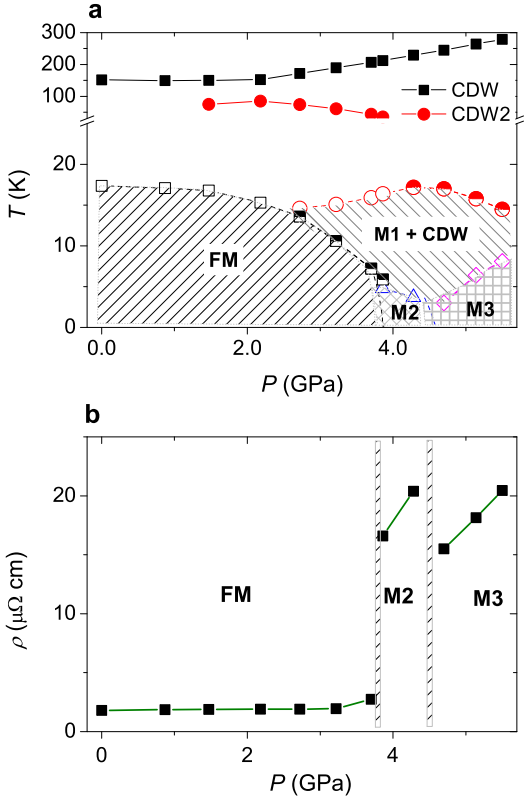}
\caption{(Color online) (a) Phase transition temperatures plotted against pressure. CDW phase transitions $T_{CDW}$ (solid squares) and $T_{CDW2}$ (solid circles) occur at high temperatures, while the FM phase transition $T_c$ (open squares) at low temperatures is intercepted by new phases $T_{M2}$ (open triangles) and $T_{M3}$ (open diamonds). Half-filled squares ($T_c$) for $P>2.14$~GPa and half-filled circles ($T_{M1}$) for $P>4.28$~GPa represent second order or weakly first order nature of the phase transitions. (b) Resistivity of SmNiC$_2$ at 1.2~K is plotted as a function of pressure. The discontinuities in $\rho$ observed around 3.8 and 4.5~GPa reflect magnetic quantum phase transitions from FM to M2 and to M3, successively.}
\end{figure}

The low-temperature resistivity of SmNiC$_2$ is magnified in Fig.~3d, which reveals contrasting behaviors across 3.8~GPa, a projected critical pressure for a FM quantum phase transition: $\rho$ is sharply reduced below $T_c$ for $P<3.8$~GPa and is enhanced below a characteristic temperature $T_{M2}$ for $P>3.8$~GPa, suggesting a gap opening on the Fermi surface. With increasing pressure, $T_{M2}$ decreases and is suppressed below 1.2~K at 4.7~GPa. A new low-$T$ phase appears at 4.7~GPa and 3.0~K, whose transition temperature $T_{M3}$ increases with pressure. Singular quantum fluctuations in the vicinity of a projected quantum critical point (QCP) have been proposed as a route to  novel quantum phases, where the novel phase essentially hides the presence of a QCP.\cite{coleman05, park06} The successive appearance of M2 and M3 phases that intercepts the FM phase may be associated with the abundant FM fluctuations near 3.8~GPa. Figure~4 summarizes the temperature-pressure phase diagram of SmNiC$_2$ and the isothermal electrical resistivity values at 1.2~K that reflect the evolution of the magnetic ground states as a function of pressure.

\begin{figure}[tbp]
\includegraphics[width=8cm]{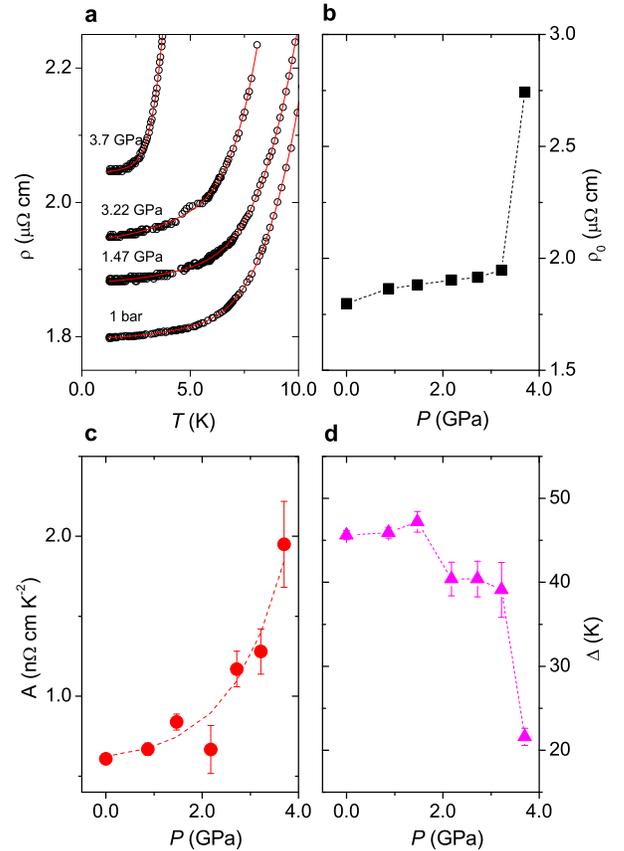}
\caption{(Color online) (a) Temperature dependence of the electrical resistivity of SmNiC$_2$ at 1~bar, 1.47, 3.22, and 3.7~GPa. Solid lines are least-squares fits of data to $\rho=\rho_{0}+AT^{2}+C_{m}T\Delta(1+2T/\Delta)$exp$(-\Delta /T)$, where the residual resistivity $\rho_0$, the $T^2$ coefficient $A$, and the magnon gap $\Delta$ are obtained from the best fits and plotted as a function of pressure in (b), (c), and~(d), respectively. Error bars shown in (c) and (d) are standard deviation from the least-squares fit. The symbols in (b) are larger than the error bars.}
\label{fig4}
\end{figure}

Figure~5a representatively shows the low-$T$ resistivity of SmNiC$_2$ within the FM phase for pressures below 3.8~GPa. For clarity, data at 3.7~GPa are rigidly shifted downward by $-0.7\mu\Omega \cdot$cm. In order to explain the temperature dependence of the resistivity below $T_c$, we use the following form that is often applied to a non-cubic ferromagnetic material:\cite{andersen79, fukuhara09} $\rho=\rho_{0}+AT^{2}+C_{m}T\Delta(1+2T/\Delta)e^{-\Delta /T}$, where the first and second terms are from impurity potential and electron-electron scattering, respectively. Scattering from magnons is represented by the third term, where $C_m$ is a constant and $\Delta$ is the magnon gap amplitude. Solid lines are from a least-squares fit of the above formula and pressure evolution of the fitting parameters $\rho_0$, $A$, and $\Delta$ is plotted in Fig.~5b, 5c, and 5d, respectively. The residual resistivity $\rho_0$ gradually increases from 1.798~$\mu\Omega\cdot$cm at 1~bar to 1.947~$\mu\Omega\cdot$cm at 3.22~GPa, then sharply increases to 2.743~$\mu\Omega\cdot$cm at 3.7~GPa. The $T^2$ coefficient $A$ exponentially increases from 0.608~$n\Omega\cdot$cm$\cdot$K$^{-2}$ at 1~bar to 1.95~$n\Omega\cdot$cm$\cdot$K$^{-2}$ at 3.7~GPa (dashed line in Fig.~4c), suggesting an enhancement of the effective mass of SmNiC$_2$. In contrast, the magnon gap $\Delta$ that characterizes the FM state decreases with pressure. The enhancement of $\rho_0$ and $A$ and suppression of the magnon gap $\Delta$ near 3.8~GPa underscores the possibility of a hidden FM QCP near 3.8~GPa.

To summarize, we have established a global phase diagram of SmNiC$_2$ by measuring electrical resistivity under pressure up to 5.5~GPa. The CDW transition temperature $T_{CDW}$ initially decreases with pressure, reaches a minimum near 1.47~GPa, then increases sharply to 279.3~K at 5.5~GPa. An additional CDW phase appears within the original CDW phase at 1.47~GPa where $T_{CDW}$ reaches a minimum, manifesting that nesting of the Fermi surface is important to formation of the CDWs. The first-order FM phase transition temperature $T_c$ is gradually suppressed with pressure, but bifurcates into the upper $M1$ and the lower FM phases above 2.18~GPa, where the lower FM transition changes to a second order or a weakly first order nature. The low-temperature resistivity was analyzed in terms of a Landau-Fermi liquid $T^2$ dependence and magnon scattering in the FM phase. The residual resistivity and the $T^2$ coefficient $A$ increase with increasing pressure, while the magnon gap is almost constant up to 3.2~GPa and is sharply suppressed at 3.7~GPa, suggesting a hidden FM quantum critical point near 3.8~GPa.

This work was supported by NRF grant funded by the Korea government (MEST) (No. 2011-0021645, \& 220-2011-1-C00014). Work at Los Alamos was performed under the auspices of the U. S. Department of Energy/Office of Science and supported in part by the Los Alamos LDRD program. VAS is supported in part from Russian Foundation for Basic Research (RFBR grant 12-02-00376). YSK acknowledges support from NRF grant funded by MEST (No. 2012K1A4A3053565). TP thanks J.~Han and J.~Shim for useful discussion.

\end{document}